# Consecutive Ones Property and PQ-Trees for Multisets: Hardness of Counting Their Orderings[*]

*Giovanni Battaglia*
Dipartimento di Informatica
Università di Pisa
gbattag@di.unipi.it

*Roberto Grossi*
Dipartimento di Informatica
Università di Pisa
grossi@di.unipi.it

*Noemi Scutellà*[†]
List SpA, Pisa
n.scutella@list-group.com

## Abstract

A binary matrix satisfies the *consecutive ones property* (C1P) if its columns can be permuted such that the 1s in each row of the resulting matrix are consecutive. Equivalently, a family of *sets* $F = \{Q_1, \ldots, Q_m\}$, where $Q_i \subseteq R$ for some universe $R$, satisfies the C1P if the symbols in $R$ can be permuted such that the elements of each set $Q_i \in F$ occur consecutively, as a contiguous segment of the permutation of $R$'s symbols. We consider the C1P version on *multisets* and prove that counting its solutions is difficult (#$\mathcal{P}$-complete). We prove completeness results also for counting the frontiers of PQ-trees, which are typically used for testing the C1P on sets, thus showing that a polynomial algorithm is unlikely to exist when dealing with multisets. We use a combinatorial approach based on parsimonious reductions from the Hamiltonian path problem, showing that the decisional version of our problems is therefore $\mathcal{NP}$-complete.

## 1 Introduction

A binary matrix $M$ of size $m \times n$ satisfies the *consecutive ones property* (C1P) if its $n$ columns can be permuted such that the 1s in each row of the resulting matrix are consecutive. An equivalent definition holds for the columns by permuting the rows. The property is often formulated in terms of sets: A family of sets $F = \{Q_1, \ldots, Q_m\}$, where each $Q_i$ is a subset of the universe of symbols $R = \{r_1, \ldots, r_n\}$, satisfies the C1P if the symbols in $R$ can be permuted such that the elements of each set $Q_i \in F$ occur consecutively as a contiguous segment of the permutation of $R$'s symbols.

For example, consider the universe $R = \{\texttt{a}, \texttt{b}, \texttt{c}, \texttt{d}, \texttt{e}\}$. The C1P is not satisfied by the family $F = \{\{\texttt{a}, \texttt{b}\}, \{\texttt{b}, \texttt{c}\}, \{\texttt{b}, \texttt{d}\}\}$, since $\texttt{b}$ can have at most two adjacent symbols in any permutation of $R$. On the other hand, the family $F = \{\{\texttt{b}, \texttt{c}\}, \{\texttt{b}, \texttt{d}\}\}$ satisfies the C1P: one feasible permutation of $R$ is $x = \texttt{eacbd}$, but not all permutations of $R$ are feasible (e.g. $y = \texttt{abcde}$ is not, because the symbols $\{\texttt{b}, \texttt{d}\}$ are not consecutive in $y$).

The C1P on sets can be formulated as a C1P problem on the binary matrix $M$ obtained by associating row $i$ with set $Q_i \in F$, and column $j$ with element $r_j \in R$. Specifically, $M_{ij} = 1$ iff $r_j \in Q_i$, as shown below for our example.

|              | a | b | c | d | e |     |              | e | a | c | b | d |
|--------------|---|---|---|---|---|-----|--------------|---|---|---|---|---|
| $\{\texttt{b},\texttt{c}\}$ | 0 | 1 | 1 | 0 | 0 |     | $\{\texttt{b},\texttt{c}\}$ | 0 | 0 | 1 | 1 | 0 |
| $\{\texttt{b},\texttt{d}\}$ | 0 | 1 | 0 | 1 | 0 |     | $\{\texttt{b},\texttt{d}\}$ | 0 | 0 | 0 | 1 | 1 |

---

[*]Research partially supported by MIUR of Italy under project AlgoDEEP prot. 2008TFBWL4.
[†]This work was done while the author was at the University of Pisa.

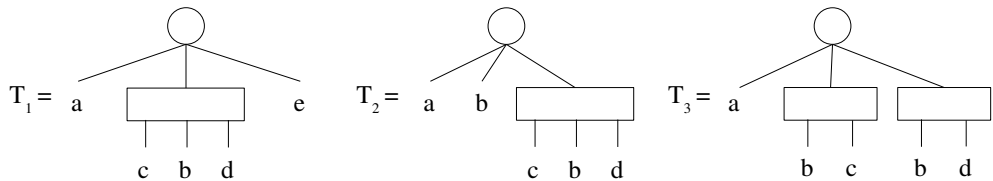

Figure 1: Some examples of PQ-trees.

The problem of finding the *orderings*, namely, the permutations of $R$ that satisfy the C1P, arises in several situations. It was first solved efficiently by Fulkerson and Gross [FG65] in their study on the incidence matrix of interval graphs, using an $O(mn^2)$ time algorithm. Ghosh [Gho72] applied the problem to information retrieval, where $R$ is the set of input records and each $Q_i$ is the set of records satisfying an answer: for each $Q_i$, the C1P guarantees that the corresponding records can be retrieved from consecutive storage locations. Booth and Leuker [Boo75, BL76] showed how to find any such ordering in linear time, with respect to the number of 1s in $M$, with applications to some graph problems such as planarity testing. They employed the *PQ-tree* data structure to represent *compactly all the orderings* yielding the C1P for the given matrix $M$.

The PQ-tree corresponding to our example is denoted by $T_1$ in Figure 1. The leaves of the PQ-tree contain the symbols of $R$: when reading these symbols by traversing the leaves in preorder, we obtain the *frontier* of the PQ-tree. As it can be seen, the frontier is one of the orderings yielding the C1P in our example tree $T_1$. Further orderings can be obtained by rearranging the children of the nodes of the PQ-tree, since they implicitly encode the sets in $F$. A round node in Figure 1 is called *P-node*, and its children can be rearranged in *any* order. A square node is called *Q-node*, and its children can be only rearranged in *left-to-right* or *right-to-left* order. By conceptually performing all the feasible rearrangements of the nodes in the PQ-tree according to the above rules, we obtain the set of frontiers for the PQ-tree. These frontiers are in *one-to-one* correspondence with all the orderings yielding the C1P for matrix $M$, as it can be verified by inspecting our example for $T_1$ (namely, $x_1 = $ acbde, $x_2 = $ adbce, $x_3 = $ aecbd, $x_4 = $ aedbc, $x_5 = $ cbdae, $x_6 = $ dbcae, $x_7 = $ cbdea, $x_8 = $ dbcea, $x_9 = $ ecbda, $x_{10} = $ edbca, $x_{11} = $ eacbd, $x_{12} = $ eadbc).

## 1.1 Our problem

Since its inception, the C1P has found many applications under several incarnations. Recent fields of application are computational biology, stringology, and bioinformatics, namely, physical mapping [JM97, COR98] and gene analysis [ELP03, LPW05, Par07], providing the inspiration for the problems in this paper. More discussion on related work is given in Section 1.3. We consider the scenario for the C1P in which the symbols in the input set $R$ are *not* necessarily distinct. We therefore investigate the problem of how to satisfy the C1P when $R$ and the $Q_i$s are *multisets*.

To get the flavor of the problem, consider the universe $R = \{a, b, b, c, d\}$ and the family $F = \{\{b, c\}, \{b, d\}\}$. The situation arises from the fact that the symbol b in both $Q_1 = \{b, c\}$ and $Q_2 = \{b, d\}$ can either match the same occurrence of b in $R$ or not. The former case gives rise to the PQ-tree $T_2$ in Figure 1, while the latter gives rise to the PQ-tree $T_3$. The set of frontiers of one PQ-tree is *not* contained in the set of frontiers of the other. However, by definition of multiset, the two occurrences of b in $R$ are *indistinguishable*. What if we impose that all the occurrences of b in $R$ must be taken simultaneously? Then, we could not deal with permutations where the occurrences of b are not contiguous, such as $y = $ abcbd, even if they satisfy the C1P. Other choices for handling multiple occurrences of b share similar drawbacks. As we will see, a polynomial-time algorithm is unlikely to exist, contrarily to what happens above for sets.

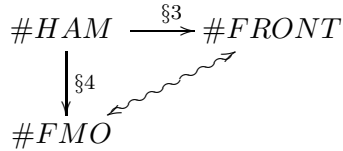

Figure 2: Relation between the counting problems described in this paper.

## 1.2 Our results

In this paper, we show that the problems dealing with the C1P on multisets are hard. Specifically, we study the problem of *counting* the number of orderings. This is "simpler" than listing all the orderings. Note that the counting problem using standard PQ-trees on *sets* takes polynomial time, since we can use the aforementioned one-to-one correspondence between the orderings and the frontiers.

The simple algorithm for sets is the following. For a given node $u$ in the PQ-tree, apply a recursive post-order traversal: If $u$ is a leaf, it has just one frontier. Otherwise, let $d$ be the number of children of $u$, and $f_i$ be the number of frontiers for the $i$th child in $u$ (where $f_i$ has been recursively computed for $1 \leq i \leq d$). Then, the number $f$ of frontiers for $u$ is $f = d! \times \prod_{i=1}^{d} f_i$ when $u$ is a P-node, and $f = 2 \times \prod_{i=1}^{d} f_i$ when $u$ is a Q-node (e.g. $f = 12 = 3! \times 2$ frontiers for $T_1$ in Figure 1).

Our first result is to prove that the problem (denoted #FRONT) of counting the frontiers of a PQ-tree whose leaves store the symbols of a *multiset* is #$\mathcal{P}$-complete [Val79]. We refer the reader to Section 3 for a proof of this result.

Turning back to the original problem (denoted #FMO) of counting the orderings for the C1P, one could hope that a polynomial solution might exist without relying on PQ-trees. This is also unlikely to happen. Our second result is to prove that the problem of counting the orderings for the C1P on multisets is #$\mathcal{P}$-complete. See Section 4 for the details of the proof.

An interesting implication of our findings is the one illustrated in Figure 2 where #HAM denotes the well-known counting version of the Hamiltonian path problem. We observed in Section 1.1 that a *direct* mapping of the orderings for the C1P in multisets into the frontiers of PQ-trees has some intrinsic ambiguity. On the other hand, we can prove that both the counting problems #FRONT and #FMO are #$\mathcal{P}$-complete using a reduction from #HAM. By the completeness properties, it follows that there exists an *indirect* mapping between the latter two problems but we do not know how to build it explicitly and directly (apart from the obvious composition).

Our approach is of independent interest, and the counting nature of the problems emphasizes their combinatorial properties. To the best of our knowledge, no previous results on C1P have been linked to either multisets or the #$\mathcal{P}$ class. In the known literature, other extensions of the C1P have been shown to be $\mathcal{NP}$-complete using reductions based on Hamiltonian paths (e.g. [Kou77]). However, our approach proves a non-weaker property of completeness (since $\mathcal{NP} \subseteq \#\mathcal{P}$) using novel reductions. Since the latter ones are *parsimonious* (i.e. preserve the number of solutions), the decisional version of #FRONT and #FMO is $\mathcal{NP}$-complete since it suffices to test non-emptiness.

## 1.3 Related work

Testing the C1P can also be done using variants of the PQ-tree data structure. Although optimal from a theoretical viewpoint, Booth and Leuker's algorithm [BL76] is quite difficult to implement since it builds the PQ-tree by induction on the number of rows of the matrix. For each row, it performs a second induction from the leaves towards the root, using one of nine templates at each node encountered in order to understand how the other nodes must be restructured.

The *PC-tree* is an alternative data structure introduced by Shih and Hsu in [SH99] to address these difficulties, that can also be used to check the C1P as shown in [Hsu01]. Both the above tree structures have remarkably simple definitions applying previously-known theorems on set families to this domain. Also, the PC-tree gives a representation of the *circular ones orderings* of the matrix $M$ just as the PQ-tree gives a representation of all the C1P orderings.

The *PQR-tree* is another alternative data structure introduced by Meidanis et al. [MPT98] to devise a tree also for the case when the input does not satisfy the C1P. In particular, the R-node is introduced: it is like the P-node, except that it captures the portion of the frontier that violates the C1P.

As previously mentioned, the C1P has several interesting applications since several apparently unrelated problems reduce to it. One of such problem is to decide if a given graph $G$ is an *interval graph*: in [FG65] the authors proved that a graph $G$ is an interval graph if and only if its *clique matrix* has the C1P by rows.

Another important application is in graph *planarity testing*: given a graph $G$ return a planar embedding for $G$ and if it does not exists return a Kuratowski subgraph isolator [Kur30]. In this case, the C1P is used as a step in the Booth and Lueker algorithm [BL76] to check planarity in linear time: this approach adds one vertex at a time, updating the PQ-tree to keep track of possible embeddings of the subgraph induced by vertices so far. (A much more simpler approach based on PC-tree has been developed in [SH99].)

Recall that not all instances of $F$ and $R$ enjoy the C1P. In that case, either duplication of symbols, or "breaking" some set in $F$ into subsets, must be allowed in order to arrange linearly the input symbols of $R$. The former scenario gives rise to the problem of minimizing duplication of symbols. The latter gives rise to the problem of minimizing the number of subsets the input sets are split into (sometimes referred in literature as the *consecutive block minimization* problem). Both problems, in their decision version, have been proved in [Kou77] to be $\mathcal{NP}$-complete (an 1.5 approximation algorithm for the block minimization problem is described in [HL08]). For example the C1P instance where $R = \{\mathtt{a}, \mathtt{b}, \mathtt{c}, \mathtt{d}\}$ and $F = \{\{\mathtt{a}, \mathtt{b}, \mathtt{c}\}, \{\mathtt{a}, \mathtt{c}, \mathtt{d}\}, \{\mathtt{b}, \mathtt{d}\}\}$ has no solution. If we allow duplication of symbols, two solutions are $x = \mathtt{bacdb}$ and $y = \mathtt{dbacd}$ (where $\mathtt{b}$ and $\mathtt{d}$ are repeated twice in $x$ and $y$ respectively). If we allow some constraints not being satisfied, an optimal solution is $z = \mathtt{bacd}$ where only the set $\{\mathtt{b}, \mathtt{d}\}$ is broken into two subsets $\{\mathtt{b}\}$ and $\{\mathtt{d}\}$.

The question of extending the C1P to multisets introduced in our paper, has not been studied before, as far as we know. It has several practical fallouts. For example, in the field of the comparative genomics, the symbols correspond to the genes, and the multisets in $F$ corresponds to set of genes occurring consecutively in one or more genomes. Genes that appear together consistently across genomes, possibly not always in the same order, are believed to be functionally related. They often code interacting proteins and have a common functional association [MPN[+]99, SLBH00, OFD[+]99].

Given two or more genomes, if each gene occurs in each genome exactly once, the above gene clusters can be modeled as *common intervals* of the permutations as described in [UY00]. These clusters can be detected by the algorithms in [UY00, HS01] in optimal linear time and space.

In [LPW05], the authors reduced the problem of finding the most "interesting" gene clusters to the C1P, with respect to the definition of maximality of [ELP03]. They employed the Booth-Lueker algorithm in order to compute the *minimal consensus* PQ-tree representing such clusters. The above paper also discusses how to handle the case where the input genomes are not permutations, but strings where some gene can be missing. For the case in which each gene can occur in a genome multiple times an exponential time algorithm is presented. Multiple occurrences of the same symbols is the way to model paralogous genes inside the same genome and multiplicity is low and rare in the observed cases [LPW05].

As seen in Section 1, if each leaf of the PQ-tree is labeled by a distinct gene symbol, it is easy to count the number of different permutations represented by the PQ-tree. This number has been used in [Par07], where the so called P-arrangements are selected among all of these permutations, to estimate the occurrence probability of a gene cluster in order to select more interesting ones. However in the same paper the problem of counting the number of different strings generated by a PQ-tree if some symbol occurs more than once is left as an open issue.

## 2  Definitions and Terminology

We consider a special class of strings defined over multisets, where the usual notions of inclusion, equality, and union, take into account the multiplicities of the elements in the multisets. We say that a string $s \equiv s_1 s_2 \cdots s_n$ is drawn from a multiset $R$ of symbols if and only if the multiset $S = \{s_1, s_2, \ldots, s_n\}$ satisfies the condition $S \subseteq R$, where $s_i$ denotes the symbol stored into position $i$ of $s$, for $1 \leq i \leq n$.

We also say that a *multiset $P$ occurs* in a *string $s$* (or equivalently $P$ is *contained* in $s$), if there is a substring $s_i s_{i+1} \cdots s_j$ of $s$, where $1 \leq i, j \leq n$, such that $P = \{s_i, s_{i+1}, \ldots, s_j\}$.[1] In the latter case, we say that $P$ occurs at position $i$ in $s$ (and $P$ is called $\pi$-pattern [AALS03]). For example, $P = \{\mathtt{a}, \mathtt{c}, \mathtt{a}\}$ occurs at position $i = 1$ in $s = \mathtt{aacb}$, while $P$ is not contained in $s_2 = \mathtt{aabc}$.

We also consider Sperner collections [Eng97] in the next sections. A collection of multisets $Q_1$, $Q_2$, ..., $Q_m \subset R$ is said to be a *Sperner Collection* (or Sperner Family, or Sperner System) if it is an anti-chain in the inclusion lattice over the powerset of $R$; namely, no multiset $Q_i$ is contained in any other multiset $Q_j$ of the collection ($i \neq j$). If no set $Q_i$ is contained in the union of the others, $\cup_{j \neq i} Q_j$, then the Sperner Collection is said to be *strict*.

Given a decision problem $\mathcal{A}$, we will denote by $\#\mathcal{A}$ its *counting version*, where we are required to count the number of the solutions of $\mathcal{A}$ [Val79]. We now introduce the #FMO problem, that formalizes the problem of extending the Booth-Leuker approach [BL76] for the C1P to multisets.

**Problem 1 (#FMO = Counting Full Multiset Orderings)** *Input: an instance $\langle R, F \rangle$, where $R$ is a multiset of symbols, and $F = \{Q_1, \ldots, Q_m\}$ is a family of multisets $Q_i \subset R$. Output: how many strings $x$ can be drawn from* all *symbols in $R$ ($|x| = |R|$), so that each $Q_i$ is contained in $x$?*

For example, given $R = \{\mathtt{a}, \mathtt{b}, \mathtt{b}, \mathtt{c}, \mathtt{d}\}$ and $F = \{\{\mathtt{b}, \mathtt{c}\}, \{\mathtt{b}, \mathtt{d}\}\}$, $x = \mathtt{abcbd}$, is one of the feasible solution of the $\langle R, F \rangle$ #FMO instance. We now introduce our second problem, which requires some additional terminology and is related to Problem 1.

**Problem 2 (#FRONT= Counting PQ-trees Frontiers)** *Input: a PQ-tree $T$, where its leaves are labeled with symbols that are not necessarily distinct. Output: what is the size of the frontier $Fr(T)$ of $T$?*

A *PQ-tree* is a tree-based data structure introduced in [Boo75, BL76] to represent succinctly a set of permutations on a set $R$ of elements, through feasible rearrangements of the children at its internal nodes. PQ-trees are useful to solve problems where the goal is to find an ordering of the input set of elements satisfying some given constraints, as in the case for the C1P.

Specifically, a PQ-tree is a rooted tree whose internal nodes are of two types: *P-nodes* that do not define any specific ordering among their children; *Q-nodes* whose children can appear either in left-to-right order or in right-to-left order. Each leaf of a PQ-tree $T$ is labeled with a symbol of the input alphabet $R$, and the *frontier* of $T$, denoted by $F(T)$, is the permutation of the symbols obtained by reading the labels of the leaves from left to right.

---
[1] In order to simplify the notation, we will always assume that an index $i$ is well defined, without explicitly writing its range when it can be deduced from the context. For example, a nonempty substring $s_i s_{i+1} \cdots s_j$ has $1 \leq i \leq j \leq n$.

Given two PQ-trees $T$ and $T'$, we say that $T$ is *equivalent* to $T'$ (written $T \equiv T'$) if one tree can be obtained from the other by possibly permuting the children of one or more P-nodes, and by possibly reversing the children of some Q-nodes. The set of the frontiers of all the trees that are equivalent to $T$ is denoted by $Fr(T)$.

Since a P-node (or a Q-node) having one child can be removed from $T$ without changing $Fr(T)$, and a P-node with two children can be replaced by a Q-node (it represents the "left to right" and "right to left" permutations only), we define the *canonical form* constraining each Q-node to have at least two children, and each P-node to have at least three children. In the rest of the paper, we assume that each PQ-tree is in canonical form.

We are interested in counting the number of frontiers in Problem 2, namely, the size of $Fr(T)$ for a PQ-tree $T$. A formal description of the $\#\mathcal{P}$ class is beyond the scope of this paper, and we refer the interested reader to the textbooks in [AB09, GJ79, Pap94]. However, we are going to use the notion of $\#\mathcal{P}$-completeness to address the difficulty of our combinatorial problems, and so we recall some basic definitions.

Let $f$ be a integral function defined over strings in $\Sigma^*$, for a given alphabet $\Sigma$. We say that $f \in \#\mathcal{P}$ if there exists a binary relation $T(-,-)$ such that:

- If $(y, x) \in T$, the length of solution $x$ is polynomial in the length of input $y$.
- It can be verified in polynomial time that a pair $(y, x)$ belongs to $T$.
- For every input $y \in \Sigma^*$, $f(y) = |\{x : (y, x) \in T\}|$ is the number of solutions for $y$.

Given two integral functions $f, g$ defined over $\Sigma^*$, we say that there exists a *polynomial Turing reduction* from $g$ to $f$ if the function $g$ can be computed in polynomial time by using a (polynomial) number of calls to an oracle for $f$. The reduction is *parsimonious* if it preserves the number of solutions.[2] A function $f$ is $\#\mathcal{P}$-*hard* if for every $g \in \#\mathcal{P}$ there is a polynomial reduction from $g$ to $f$. As usual, a function is $\#\mathcal{P}$-*complete* if it is both $\#\mathcal{P}$-hard and it is in $\#\mathcal{P}$.

## 3 Counting the frontiers of a PQ-tree

We begin by discussing the completeness of the #FRONT problem. We use a reduction from the well-known counting version of Hamiltonian Path (#HAM). We are given an undirected graph $G$, a source vertex $w \in G$, and a destination vertex $s \in G$. We want to know how many paths $H$ in $G$ start in $w$ and end in $s$, such that all the vertices in $G$ are traversed exactly once by each $H$. For example, one such path is $H = \langle 1, 3, 2, 4, 5 \rangle$ in the graph $G$ shown in Figure 3. In the rest of the paper, we assume that $G$ is connected, $w$ and $s$ have degree at least one, and the other vertices have degree at least two (otherwise there is no Hamiltonian path). We also assume that there are no multiple edges between the same pair of vertices and no self-loops.

### 3.1 Construction of the PQ-trees

The main idea is to code the structure of the given graph $G$ in three suitable PQ-trees, $T_G$, $T_V$, and $T_E$, such that each Hamiltonian path $H$ is in one-to-many correspondence with a suitable set of strings from their frontiers. We now describe our reduction from $G = \langle V, E \rangle$ to $T_G$, $T_V$, and $T_E$, using Figure 3 as an illustrative example.

---
[2] Hence it allows for non-emptiness testing in the decisional version of the problems.

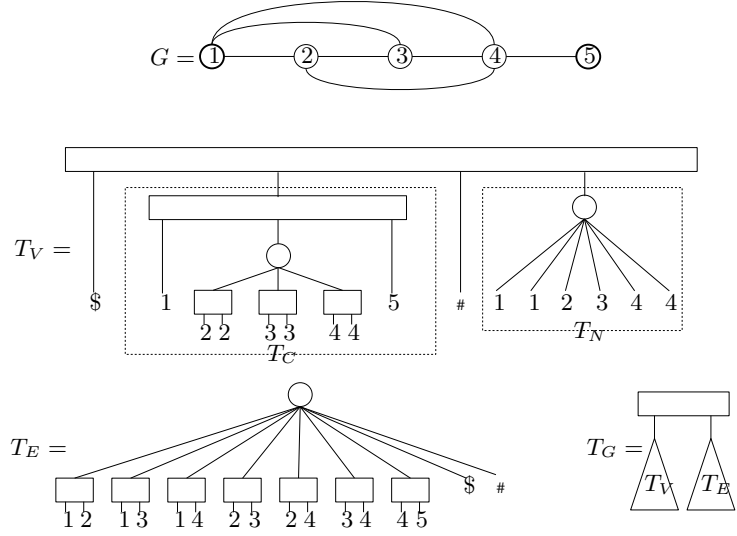

Figure 3: The PQ-tree $T_G$ associated with the input graph $G$ where the source and the destination vertices are $w = 1$ and $s = 5$. Note that $T_G$ has $T_V$ and $T_E$ are shown individually.

The root of $T_G$ is a Q-node having two PQ-trees $T_V$ and $T_E$ as children.

Tree $T_E$ encodes all the feasible permutations of the edges in $E$. The root of $T_E$ is a P-node having $|E|+2$ children. Two of them are special "endmarkers," and are labeled with $\$$ and $\#$. Each of the remaining children is a Q-node that encodes an edge $e = \{i, j\}$ by two leaves labeled with $i$ and $j$, respectively, as children. In our example, $T_E$ has $|E| = 7$ Q-nodes with children labeled by $\{1,2\}$, $\{1,3\}$, $\{1,4\}$, $\{2,3\}$, $\{2,4\}$, $\{3,4\}$, and $\{4,5\}$, plus the endmarkers $\$$ and $\#$.

Tree $T_V$ enforces a classification of the edges as "coding" a Hamiltonian path, or "non-coding" otherwise. Specifically, the root of $T_V$ is a Q-node with four children: one leaf labeled with $\$$, a PQ-tree $T_C$ for the coding edges, one more leaf labeled with $\#$, and a PQ-tree $T_N$ for the non-coding edges. The root of $T_C$ is a Q-node with three children. The first child is a leaf labeled with the source $w$ and the last is a leaf labeled with the destination $s$. The middle child is a P-node with $|V| - 2$ children, each of which is a Q-node with two leaves labeled with the same symbol $i$, for $i \in V \setminus \{w, s\}$. In our example $w = 1$, $s = 5$, and $|V| = 5$. The root of the non-coding tree $T_N$ is a P-node having $2(|E| - |V| + 1)$ leaves as children. Letting $d_i$ denote the degree of vertex $i$, there are $d_w - 1$ leaves labeled with $w$, $d_s - 1$ leaves labeled with $s$, and $d_i - 2$ leaves labeled with $i \neq w, s$. In our example, the leaves are labeled with $1, 1, 2, 3, 4, 4$, where $2(|E| - |V| + 1) = 6$.

The above construction requires polynomial time, and the rationale will be given in Section 3.2.

**Lemma 1** *Given a undirected graph $G = \langle V, E \rangle$, its corresponding PQ-trees $T_G$, $T_V$, and $T_E$ can be built in $\mathcal{O}(|V| + |E|)$ time.*

## 3.2 Properties of the PQ-trees

Consider the Hamiltonian path $H = \langle 1, 3, 2, 4, 5 \rangle$ in our example. (Observe that the reversal of $H$, namely $\langle 5, 4, 2, 3, 1 \rangle$, is also a Hamiltonian path, but we consider it to be different from $H$ for the counting purposes.) The corresponding strings $\alpha^H$ belonging to the frontiers $Fr(T_G)$ are thus characterized. First at all, each $\alpha^H$ is a square, namely, the concatenation $\alpha^H = \alpha \alpha$ of two equal strings $\alpha$, where $\alpha$ belongs to both the frontiers $Fr(T_V)$ and $Fr(T_E)$, and is of length $2|E| + 2$. For example, $\alpha = \$13322445\#121434$ is one such feasible string. We can characterize the general structure of the strings $\alpha$ by observing that they matches one of the following two

patterns. Let $\pi$ denote an arbitrarily chosen permutation of the pairs in $\{1,2\}, \{1,4\}, \{3,4\}$, which represent the edges *not* traversed by $H$. (That is, $\pi$ belongs to the frontiers of the PQ-tree resulting from $\{\{1,2\}, \{1,4\}, \{3,4\}\}$.) The former pattern for $\alpha$ is $\$\,13322445\,\#\,\pi$, where the initial symbols are fixed and only $\pi$ may vary; analogously, the latter is $\pi\,\#\,13322445\,\$$. For example, $\alpha = 413421\,\#\,13322445\,\$$ matches the latter pattern.

Having introduced the structure of $\alpha^H = \alpha\alpha$ in our example, we show how to make $\alpha$ satisfy the implicit conditions encoded in $T_V$ and $T_E$. Indeed, $T_E$ guarantees that the two integers in each of the pairs corresponding to the edges in $E$ always occur consecutively in $\alpha$. Moreover, the subtree $T_C$ in $T_V$ constraints each vertex $i \in V \setminus \{w, s\}$ to appear exactly twice in the chosen subset of edges, while $w$ and $s$ are required to appear just once. Note that the purpose of the subtree $T_N$ is that of "padding" the edges in $E$ that are not traversed by $H$, since we do not know a priori which ones will be touched by $H$.

We now generalize the above observations on $\alpha$. In the following we can restrict our focus on paths of the form $i_1, i_2, \ldots, i_{|V|}$, that are permutations of $\{1, 2, \ldots, |V|\}$ with $i_1 = w$ and $i_{|V|} = s$ (otherwise they cannot be Hamiltonian paths from $w$ to $s$). Moreover we introduce the notation $Perm(Q)$ for a set $Q = \{\{a_1, b_1\}, \{a_2, b_2\}, \ldots, \{a_r, b_r\}\}$ of unordered pairs. It represents the set of all the permutations of $a_1, b_1, a_2, b_2, \ldots, a_r, b_r$ such that $a_l$ and $b_l$ occupy contiguous positions for $1 \le l \le r$. For example, given $Q = \{\{1,2\}, \{1,4\}, \{3,4\}\}$, we have that 413421 is a valid permutation in $Perm(Q)$, while 413241 is not.

We now show in Lemmas 2–4 that there exists a one-to-many correspondence between the Hamiltonian path $H$ in $G$ and the strings $\alpha \in Fr(T_V) \cap Fr(T_E)$.

**Lemma 2** *Let $G = \langle V, E \rangle$ be an undirected graph, and $T_G$, $T_V$, and $T_E$ be its corresponding PQ-trees. For any string $\alpha \in Fr(T_V) \cap Fr(T_E)$, there exists a corresponding Hamiltonian path $H$ of $G$ from $w$ to $s$.*

*Proof*: Consider a string $\alpha \in Fr(T_V) \cap Fr(T_E)$. We first show that the symbols in $\alpha$ follow a special pattern.

Since $\alpha \in Fr(T_V)$, the symbols \$ and \# in it match those in the leaves of $T_V$ by construction. Assume w.l.o.g. that the first symbol of $\alpha$ is \$. (The other case in which \$ is the last symbol of $\alpha$ is analogous.) Then, $\alpha$ is of the form $\alpha = \$\,\tau\,\#\,\pi$ by construction, where $\tau = \tau_1 \tau_2 \cdots \tau_{2|V|-2}$ and $\pi$ should follow the patterns described next. First, $\tau = w\tau's$ where $\tau' \in Perm(\{i,i\}_{i \ne w,s})$, since $\tau' \in Fr(T_C)$: hence, $\tau_i = \tau_{i+1}$ for even values of $i \in [2 \ldots 2|V|-4]$. Second, $\pi$ is a permutation of the symbols in the multiset obtained by removing the symbols of $\tau$ from $\bigcup_{\{i,j\} \in E}\{i,j\}$.

Now, the fact that $\alpha$ belongs also to $Fr(T_E)$ puts additional constraints on $\tau$ and $\pi$. Indeed, the Q-nodes in $T_E$ guarantee that $\tau_1$ and $\tau_2$ are children of the same Q-node, $\tau_3$ and $\tau_4$ are children of the next Q-node, and so on. Thus in general $\tau_i, \tau_i + 1$ for odd $i$ belong to the same Q-node: hence, $\{\tau_i, \tau_i + 1\} \in E$, for even values of $i \in [2 \ldots 2|V|-4]$. Combining the latter with the fact that $\tau_i = \tau_{i+1}$ for odd values of $i$, we obtain that $H = \langle w, \tau_2, \ldots, \tau_{2|V|-4}, s \rangle$ is a Hamiltonian path.

Note that the rest of the Q-nodes in $T_E$ induce also some contiguity constraints on $\pi$, which will be relevant later for the counting argument (see Lemma 5). The case $\alpha = \pi\,\#\,\tau\,\$$ is analogous. □

**Lemma 3** *Let $G = \langle V, E \rangle$ be an undirected graph, and $T_G$, $T_V$, and $T_E$ be its corresponding PQ-trees. For any Hamiltonian path $H$ of $G$ from $w$ to $s$, there exists at least one corresponding string $\alpha \in Fr(T_V) \cap Fr(T_E)$.*

*Proof*: Let $H = \langle i_1, i_2, \ldots, i_{|V|} \rangle$ be a Hamiltonian path, where $i_1 = w$ and $i_{|V|} = s$. We define $\alpha = \$\,\tau\,\#\,\pi$ where $\tau$ and $\pi$ are as follows. First, we choose $\tau = i_1 i_2 i_2 \cdots i_{|V|-1} i_{|V|-1} i_{|V|}$, so that

$\tau \in Fr(T_C)$. Second, let $E' = E \setminus \{\{i_j, i_{j+1}\}\}_{1 \leq j \leq |V|-1}$ be the set of edges not traversed by $H$. Let list the edges of $E'$ as $\{a_1, b_1\}, \ldots, \{a_r, b_r\}$. Then we choose $\pi = a_1 b_1 \cdots a_r b_r$, so that $\pi \in Fr(T_N)$.

Consequently, $\alpha$ should belong to $Fr(T_V)$. It remains to see that $\alpha$ belongs also to $Fr(T_E)$. Note that the $\$$ and $\#$ symbols in $\alpha$ clearly match the two endmarker leaves in $T_E$. Also, by our construction of $\tau$ and $\pi$, for any edge $\{i, j\}$ in $E$, we have that $i$ and $j$ appear in consecutive positions of either $\tau$ or $\pi$. This concludes the proof implying that $\alpha \in Fr(T_V) \cap Fr(T_E)$. □

**Lemma 4** *Let $\Sigma_H \subseteq Fr(T_V) \cap Fr(T_E)$ denote the set of all the strings corresponding to a given Hamiltonian path $H$, as stated in Lemma 3. Then, for any two Hamiltonian paths $H \neq H'$ of $G$ from vertex $w$ to vertex $s$, it is $\Sigma_H \cap \Sigma_{H'} = \emptyset$.*

*Proof*: For any $\alpha \in \Sigma_H$ and $\alpha' \in \Sigma_{H'}$, we show that $\alpha \neq \alpha'$. If one of the strings begins with the $\$$ symbol, while the other does not, they are different since neither $\tau$ or $\pi$ contains any endmarker (e.g. $\alpha = \$\tau\#\pi$ is different from $\alpha' = \pi'\#\tau'\$$). Hence, consider the case when both $\alpha$ and $\alpha'$ begin with $\$$. Since the corresponding Hamiltonian paths $H$ and $H'$ are different, also the corresponding "coding" strings $\tau$ and $\tau'$ will be different by construction, implying that $\alpha \neq \alpha'$. □

### 3.3 Reduction from #HAM to #FRONT

We now show how to reduce the problem #HAM of counting the Hamiltonian paths in $G = \langle V, E \rangle$, to the problem #FRONT of counting the frontiers of PQ-trees, namely, $T_G$, $T_V$, and $T_E$. We denote the number of frontiers for a PQ-tree $T$ by $|Fr(T)|$. Here is the polynomial time reduction for the input graph $G$ and its two vertices $w$ and $s$:

- Build the PQ-trees $T_G$, $T_V$, and $T_E$ (see Lemma 1).

- Return the following integer as the number of Hamiltonian paths from $w$ to $s$ in $G$:

$$\frac{2\,|Fr(T_V)| \times |Fr(T_E)| - |Fr(T_G)|}{2\,(|E| - |V| + 1)! \times 2^{|E|-|V|+1}} \tag{1}$$

Clearly, the formula in (1) can be computed in polynomial time. We now show its correctness.

**Lemma 5** *Let $\Sigma_H \subseteq Fr(T_V) \cap Fr(T_E)$ denote the set of strings corresponding to a Hamiltonian path $H$. Then, for any Hamiltonian path $H$ from $w$ to $s$, we have $|\Sigma_H| = 2\,(|E| - |V| + 1)! \times 2^{|E|-|V|+1}$.*

*Proof*: Consider a string $\alpha \in \Sigma_H$. As previously mentioned in the proof of Lemma 2, $\alpha$ matches either the pattern $\$\tau\#\pi$ or $\pi\#\tau\$$. Note that the string $\tau$ is uniquely determined by construction of $T_C$, and the contiguity condition imposed by $T_E$, for the given $H$. Hence, $|\Sigma_H|$ is twice the number of strings $\pi$ that we can obtain from $T_N$, under the contiguity condition imposed by $T_E$. Therefore, $|\Sigma_H| = 2\,|Perm(E')|$, where $E' \subseteq E$ is the set of edges not traversed by $H$. Since $|E'|$ is $p = |E| - |V| + 1$, we have $p!$ permutations of these edges and, for each of them, we have two ways to permute every $\{i, j\} \in E'$. This gives a total of $p!\,2^p$ strings $\pi$. Note that we cannot generate twice the same string in this way, because the edges are distinct as unordered pairs and, for each pair $\{i, j\} \in E'$, it is $i \neq j$. Hence the result follows. □

**Lemma 6** $|Fr(T_G)| = 2\,|Fr(T_V)| \times |Fr(T_E)| - |Fr(T_V) \cap Fr(T_E)|$

*Proof*: Let $L_V = Fr(T_V)$, $L_E = Fr(T_E)$, and $L_G = Fr(T_G)$. Consider $L_{VE} = Fr(T_V) \cap Fr(T_E)$, so that we can rewrite $L_V = L'_V \cup L_{VE}$ and $L_E = L'_E \cup L_{VE}$. Now, by construction of $T_G$, we

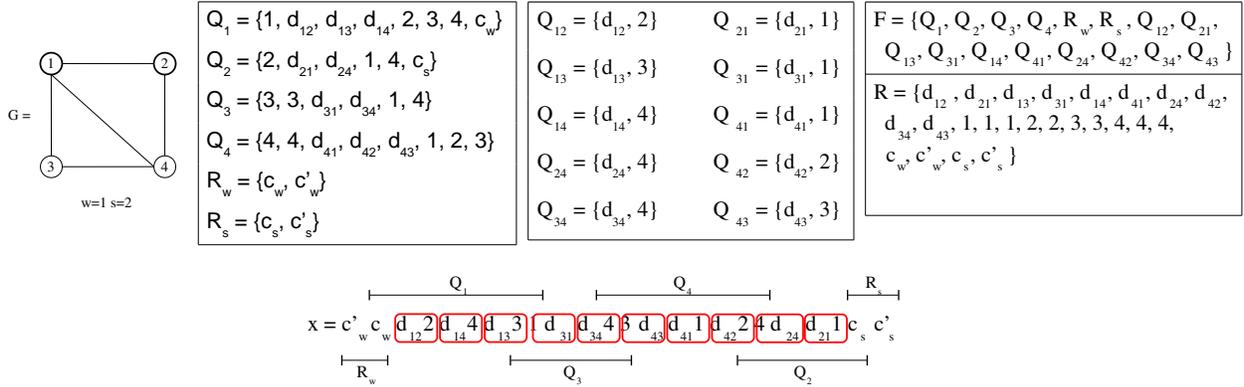

Figure 4: Example of reduction of a Hamiltonian Path instance where the source and the destination vertices are $w = 1$ and $s = 2$ into a #FMO instance $\langle R, F \rangle$. Sets $Q_{ij}$ are shown boxed in string $x$.

know that $L_G = L_V \cdot L_E \cup L_E \cdot L_V$, where the standard operation "$\cdot$" denotes the extension of the string concatenation to sets of strings (i.e. $A \cdot B = \{ab \mid a \in A, b \in B\}$). By expanding $L_V$ and $L_E$, we obtain that $L_G = (L'_V \cup L_{VE}) \cdot (L'_E \cup L_{VE}) \cup (L'_E \cup L_{VE}) \cdot (L'_V \cup L_{VE})$. By simple algebra, we have that $|L_G| = |L_V \cdot L_E| + |L_E \cdot L_V| - |L_E \cap L_V|$. The result follows, since $|L_V \cdot L_E| = |L_E \cdot L_V| = |L_E| \times |L_V|$. $\square$

We now have all the ingredients to prove the #$\mathcal{P}$-completeness of the #FRONT problem.

**Theorem 7** *#FRONT is #$\mathcal{P}$-complete.*

*Proof*: The membership to #$\mathcal{P}$ trivially holds. In order to prove that the formula in (1) is correct, observe that the sets $\Sigma_H$ for all the Hamiltonian paths $H$ from $w$ to $s$, are a partition of $I = Fr(T_V) \cap Fr(T_E)$. To see why, note that for each string in $I$, there is a Hamiltonian path by Lemma 2. Moreover, $\Sigma_H \subseteq I$ by Lemma 3. Finally, the sets $\Sigma_H$ are pairwise disjoint by Lemma 4.

Formula (1) is based on the fact that $|I|$ can be obtained from $|T_G|$, $|T_V|$, and $|T_E|$ by using Lemma 6. Moreover, sets $\Sigma_H$ have all the same size, as stated in Lemma 5. Hence, dividing these two quantities gives an integer as a result, which is the number of Hamiltonian paths as in (1). Note that our reduction requires polynomial time. $\square$

## 4 Hardness results for #FMO

We now show how to reduce the #HAM problem to the counting version of the Full Multiset Problem (#FMO). For the given undirected graph $G = \langle V, E \rangle$, together with the source and the destination vertices, $w$ and $s$, we make the same assumptions as in Section 3. In Section 4.1, we walk through the example in Figure 4 to describe the reduction. In Section 4.2, we characterize the structure of each string satisfying the constraints in the #FMO instance. In Section 4.3, we prove our hardness result on counting how many strings correspond to the same Hamiltonian path $H$ in $G$.

### 4.1 Instance construction

Consider the example in Figure 4. On the left we show the input undirected graph $G$, where the source and the destination vertices $w = 1$ and $s = 2$ are in boldface. The corresponding #FMO instance $\langle R, F \rangle$ is reported on the right, while one of the solution string $x$, corresponding to the Hamiltonian path $H = \langle 1, 3, 4, 2 \rangle$ is represented at the bottom.

We build an instance of #FMO as follows. For each vertex $i$, we construct the multiset $Q_i$ containing two occurrences of the symbol $i$ (if $i \neq w, s$), or one occurrence of $i$ and one of the special symbol $c_i$ (if $i = w, s$). We also add symbols $d_{ij}$ and $j$ to $Q_i$, for every incident edge $\{i, j\}$. As a result, each undirected edge $\{i, j\}$ is represented by two different symbols $d_{ij} \in Q_i$ and $d_{ji} \in Q_j$. Formally,

$$Q_i = \begin{cases} \bigcup_{\{i,j\} \in E} \{d_{ij}, j\} \cup \{i, c_i\} & i = w, s \\ \bigcup_{\{i,j\} \in E} \{d_{ij}, j\} \cup \{i, i\}, & i \neq w, s \end{cases}$$

To guarantee the condition that $w$ and $s$ are the source and the destination vertex, we introduce two symbols $c'_w$ and $c'_s$, and two sets $R_w = \{c_w, c'_w\}$ and $R_s = \{c_s, c'_s\}$, which do not correspond to any vertex of the input graph. They are used to guarantee that $Q_w$ and $Q_s$ will always occur as the first and the last multiset of any solution string $x$ for our #FMO instance.

In general, the intersection between two multisets $Q_i$ and $Q_j$ can contain more symbols than just $i$ and $j$. For example, the intersection between $Q_1$ and $Q_4$ is $I_{14} = \{1, 4, 2, 3\}$ because it contains also 2 and 3, each of them corresponding to the vertex forming a triangle with 1 and 4. To avoid this situation, $2|E|$ auxiliary multisets $Q_{ij} = \{d_{ij}, j\}$ are used to constraint the intersection between the multisets inside each solution string $x$, such that it contains exactly two symbols. Observe that each edge $\{i, j\} \in E$ gives rise to two multisets $Q_{ij}$ and $Q_{ji}$. In the string $x$ shown in Figure 4, the purpose of the multisets $Q_{ij}$ and $Q_{ji}$ is to enforce the intersection between $Q_1$ and $Q_3$ inside $x$ to be $\{1, 3\}$, between $Q_3$ and $Q_4$ to be $\{3, 4\}$, and so on.

We finally choose the multiset $R = Q \setminus R'$ where $Q = \bigcup_i Q_i \cup \{c'_w, c'_s\}$ and $R' = \bigcup_{i \neq w, s} \{i, i\} \cup \{w, s\}$. We also choose $F = \{Q_1, \ldots, Q_{|V|}\} \cup \{R_w, R_s\} \cup \{Q_{ij}, Q_{ji}\}_{\{i,j\} \in E}$. The idea behind the construction of $R$ and $F$ is illustrated in our example. Each Hamiltonian path $H$ from $w = 1$ to $s = 2$ contains only one edge incident to $w$ ($\{1, 3\}$ in our example), one edge incident to $s$ ($\{2, 4\}$), and two edges incident to each of the other vertices in $H$ ($\{1, 3\}$ and $\{3, 4\}$ incident to 3, and $\{3, 4\}$ and $\{2, 4\}$ incident to 4). The path $H$ can always be represented by a string $x$ having size $|R|$. The multisets $Q_i$ occur inside $x$ in the same order as that of the vertices $i$ inside $H$. The intersection between consecutive $Q_i$ and $Q_j$ is now guaranteed to contain just $i, j$ in consecutive positions of $x$. For example, $Q_1, Q_3, Q_4,$ and $Q_2$ correspond to the vertices in $H = \langle 1, 3, 4, 2 \rangle$, while their intersections correspond to the edges used in $H$. Here is the role of $R'$: since we do not know a priori which edges will be traversed by $H$, we can rely just on the multiset given by their endpoints, thus giving rise to $R'$. Even if we have to remove $R'$ from $Q$ to obtain $R$, we still guarantee that $\langle R, F \rangle$ is a valid #FMO instance.

**Lemma 8** *Each multiset $M \in F$ is contained in $R$.*

*Proof*: We recall that $F = \{Q_1, \ldots, Q_{|V|}\} \cup \{R_w, R_s\} \cup \{Q_{ij}, Q_{ji}\}_{\{i,j\} \in E}$, and that $R = Q \setminus R'$ where $Q = \bigcup_i Q_i \cup \{c'_w, c'_s\}$ and $R' = \bigcup_{i \neq w, s} \{i, i\} \cup \{w, s\}$. Since we assumed that the degree of $w$ is at least one, $w$ has at least one incident edge $\{w, j\}$. By construction of the $Q_i$ multisets, it follows that the symbol $w$ has at least two occurrences in $Q$: one occurrence belongs to $Q_w$, while the second occurrence belongs to the multiset $Q_j$ associated to the vertex $j$. Same as above for the destination vertex $s$, that occurs at least two times in $Q$. Since we assumed each one of the remaining vertex $i \neq w, s$, to have at least two neighbors in $G$, (let say $j, l$,) it follows that the symbol $i$ has at least four occurrences in $Q$: two occurrences belong to $Q_i$, the third occurrence belongs to $Q_j$, while the fourth one belongs to $Q_l$.

From the above, it follows that $R = Q \setminus R'$ contains at least one occurrence of $w$, one occurrence of $s$, and two occurrence of each $i \neq w, s$.

At this point, we have all the ingredients to prove that $Q_w \subseteq R$. The multiset $Q_w$ contains exactly one occurrence of $w$, and at most one occurrence for every other symbol $i \neq w$. Moreover,

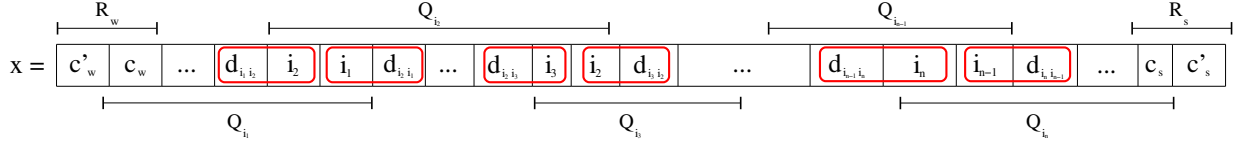

Figure 5: The string $x$ coding the Hamiltonian path $H = \langle i_1, \ldots, i_n\rangle$ of $G$. Intersections between $Q_i$ and $Q_j$ have size 2 in $x$ and are constrained to be $\{i, j\}$.

for each $d_{ij} \in Q_w$, it holds that $d_{ij} \in R$, since $R \subseteq Q$, and no one $d_{ij}$ is in $R'$. Also the symbol $c_w$ is contained in $R$, since $c_w \in Q_w$, but $c_w \notin R'$. Same as above for $Q_s$, and the remaining $Q_i$ multisets, with $i \neq w, s$. In the case of the $Q_i$ multisets, the symbol $i$ occurs two times inside each $Q_i$, but this is not an issue since, as discussed above, $R$ contains at least two occurrence of each symbols $i \neq w, s$.

To prove that each $Q_{ij} = \{d_{ij}, j\}$ and $Q_ij = \{d_{ji}, i\}$ is contained in $R$, it is enough to note that $d_{ij}, d_{ji} \in Q$, but $d_{ij}, d_{ji} \notin R'$, and that for every symbol $i$ or $j$ there is at least one occurrence in $R$.

Finally, we observe that $R_w = \{c_w, c'_w\}$ and $R_s = \{c_s, c'_s\}$ are contained in $R$, since the symbols $c_w, c_s, c'_w, c'_s$ are in $Q$, but they are not in $R'$. □

**Lemma 9** *Given a undirected graph $G = \langle V, E\rangle$, together with a source and a destination vertex, $w$ and $s$, the corresponding instance $\langle R, F\rangle$ of #FMO, can be built in $\mathcal{O}(|V| + |E|)$ time.*

## 4.2 Characterization of the solutions

We need some technical lemmas, as in Section 3.2. In particular, Lemmas 10–12 follow the same path as that traced in Lemmas 2–4 for #FRONT.

**Lemma 10** *Let $G = \langle V, E\rangle$ be an undirected graph, and $\langle R, F\rangle$ be its corresponding #FMO instance. For any string $x$ that is solution of $\langle R, F\rangle$, there exists a corresponding Hamiltonian path $H$ of $G$ from $w$ to $s$.*

**Lemma 11** *Let $G = \langle V, E\rangle$ be an undirected graph, and $\langle R, F\rangle$ be its corresponding #FMO instance. For any Hamiltonian path $H$ of $G$ from $w$ to $s$, there exists at least one corresponding solution $x$ of $\langle R, F\rangle$.*

**Lemma 12** *Let $\Sigma_H$ denote the set of all the solutions of $\langle R, F\rangle$ corresponding to a given Hamiltonian path $H$, as stated in Lemma 11. Then, for any two Hamiltonian paths $H \neq H'$ of $G$ from vertex $w$ to vertex $s$, it is $\Sigma_H \cap \Sigma_{H'} = \varnothing$.*

We now prove Lemma 10, leaving the proof of Lemmas 11–12 at the end of the section. We consider a solution $x$ of $\langle R, F\rangle$, and make three conceptual steps.

(a) We prove that the multisets $Q_i$ follow a total order $\prec_x$ induced by $x$.

(b) We show that each $Q_i$ occurs exactly once in $x$.

(c) For any two consecutive $Q_i$ and $Q_j$ in the total order $\prec_x$, we demonstrate that their intersection in $x$ corresponds to edge $\{i, j\} \in E$.

Observe that steps $(a)$ and $(b)$ encode all possible permutations of the vertices in $V$, while step $(c)$ selects only those permutations (if any) that correspond to paths in $G$. Putting $(a)$–$(c)$

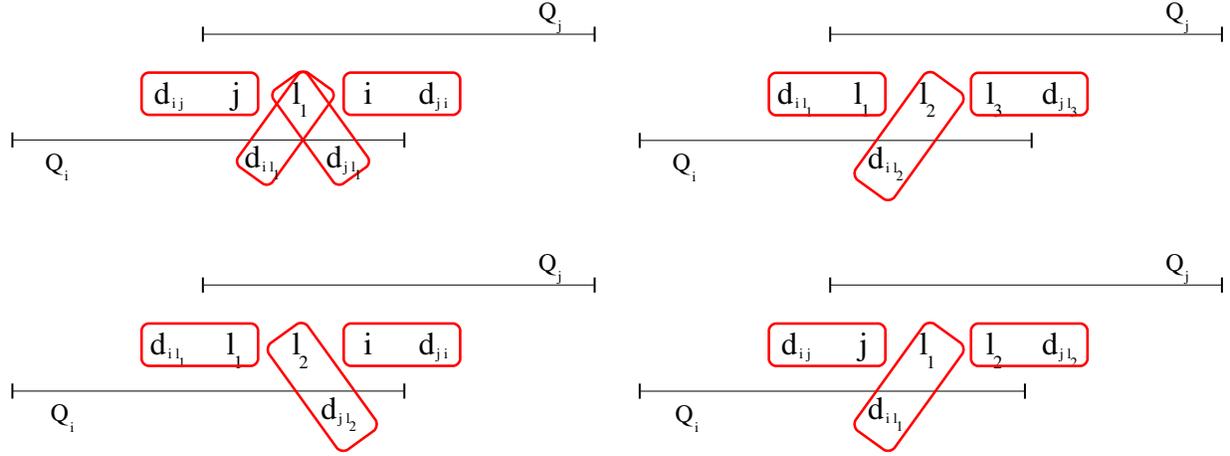

Figure 6: The four possible cases if $I_{ij} = 3$. From left to right, top-down, the case where $I_{ij} = \{i, j, l_1\}$, $I_{ij} = \{l_1, l_2, l_3\}$, $I_{ij} = \{i, l_1, l_2\}$, or $I_{ij} = \{j, l_1, l_2\}$.

together, we can see that the Hamiltonian path corresponding to $x$ is $H = \langle i_1, i_2 \ldots, i_{|V|} \rangle$, where $Q_{i_1} \prec_x Q_{i_2} \prec_x \cdots \prec_x Q_{i_{|V|}}$ is the total order induced by $x$.

We show a slightly more general property than that stated in $(a)$, using the following lemma.

**Lemma 13 (Strict Sperner Property)** *The collection of multisets $C = \{R_w, R_s, Q_1, \ldots, Q_{|V|}\}$, is a Strict Sperner collection: no multiset is contained in the union of the others. Hence, there exists a total order $\prec_x$ on the multisets in $C$.*

*Proof*: First at all we observe that each $Q_i \in C$ contains at least one symbol $d_{ij}$ that is unique in $x$ and does not belong to any other multiset. Hence, $Q_i$ cannot be contained in the union of the other multisets. Also the multisets $R_w$ and $R_s$ contain unique symbols, namely, $c'_w$ and $c'_s$. Hence, $C$ is a strict Sperner collection: this property, combined with the fact that each multiset in $C$ occurs in $x$, implies that a left-to-right scan of $x$ provides a total order of the multisets in $C$. That is, for any pair $Q_i$ and $Q_j$ either $Q_i \prec_x Q_j$ or $Q_j \prec_x Q_i$. □

We prove the property stated in step $(c)$ by the following lemma.

**Lemma 14 (Intersection Size)** *Let $x$ be a string of size $|R|$, drawn from all the symbols in $R$, and containing all the multisets in $C_2 = \{R_w, R_s, Q_i, Q_{ij}\}$. Let $I_{ij} = Q_i \cap Q_j$ denote the intersection between two multisets $Q_i$ and $Q_j$ that occur consecutively in $x$. Then, $(i)$ $|I_{ij}| = 2$; $(ii)$ $I_{ij} = \{i, j\}$; $(iii)$ $\{i, j\} \in E$.*

*Proof*: $(i)$ First, let $l_1, l_2, \ldots$ denote some generic vertices that are adjacent to both $i$ and $j$. By construction of the multisets $Q_i$, note that $I_{ij}$ can only contain the symbols $i$, $j$ or $l_p$ for $p = 1, 2, \ldots$. Formally:

$$Q_i \cap Q_j = \begin{cases} \{i,j\} \cup \bigcup_{\{i,l_p\},\{l_p,j\} \in E} \{l_p\} & \{i,j\} \in E \\ \bigcup_{\{i,l_p\},\{l_p,j\} \in E} \{l_p\} & \text{otherwise} \end{cases} \quad (2)$$

Assume that $|I_{ij}| = 3$. Then, four cases are possible when considering the sets $Q_{fg}$ where $f, g \in \{i, j, l_1, l_2, \ldots\}$ and $f \neq g$:

1. $I_{ij} = \{i, j, l_1\}$

2. $I_{ij} = \{i, l_1, l_2\}$

3. $I_{ij} = \{j, l_1, l_2\}$

4. $I_{ij} = \{l_1, l_2, l_3\}$

We discuss case 1 (since cases 2–4 are similar), which is represented on the top left of Figure 6. Here, it is shown that the symbols in the four multiset $Q_{ij}$, $Q_{ji}$, $Q_{il_1}$ and $Q_{jl_1}$, corresponding to the three edges $\{i, j\}$, $\{i, l_1\}$ and $\{j, l_1\}$, cannot occur inside $Q_i$ or $Q_j$, because each symbol $d_{ij}$ only belongs to $Q_i$ (hence it cannot be a member of the intersection $I_{ij}$), and we only have one occurrence of $l_1$ inside $Q_i$ and one occurrence inside $Q_j$.

The cases where the intersection has size larger than 3 are similar. In these cases we can always select from $I_{ij}$ a subset of three symbols, reducing to one of the above cases: if $|I_{ij}| > 3$, we can apply the above argument to $i$, $j$ and an arbitrary vertex in $I_{ij} \setminus \{i, j\}$.

Given the above upper bound on the size of an intersection, we now prove that $|I_{ij}|$ cannot be smaller than 2. By Lemma 13 we know that each multiset $Q_i$ cannot be contained in the union of the other multisets, hence in order to construct a string $x$ of size $|R|$ containing all the multisets in $C_2$, the combined size of the intersections between the $Q_i$ multisets must be $2(|V|-1)$. Assuming that at least one of such intersections has size 1, then some other intersection would have size 3, contradicting the previous upper bound. From the previous upper and lower bounds it follows that each intersection must have size $|I_{ij}| = 2$.

$(ii)$ To prove that $I_{ij} = \{i, j\}$, let us assume by contradiction that $I_{ij} = \{i, l\}$, where $l \neq j$ is a vertex forming a triangle in the input graph $G$ together with $i$ and $j$. As in point $(i)$, it is easy to prove that the two sets $Q_{il} = \{d_{il}, l\}$, $Q_{jl} = \{d_{jl}, l\}$ cannot occur inside the solution string $x$, since $Q_j$ and $Q_i$ only contain one occurrence of the symbol $l$ each. The $d_{jl}$ symbol cannot be contained in the intersection $I_{ij}$ since only the symbols $i$ and $l$ are inside.

The proofs for the other cases $I_{ij} = \{j, l_1\}$ and $I_{ij} = \{l_1, l_2\}$ are identical to this one.

$(iii)$ The conclusion follows from the point $(ii)$ and from the intersection property highlighted in Equation (2), stating that if $\{i, j\} \subseteq I_{ij}$, then $\{i, j\} \in E$. □

Finally, the property stated in step $(b)$ is based on the lemma below.

**Lemma 15 (Occurrence Uniqueness)** *Given a solution $x$ of $\langle R, F \rangle$, each multiset $Q_i \in F$ occurs exactly once inside $x$.*

*Proof*: We recall that each $Q_i$ occurs at least once inside $x$ since the latter is a valid solution. Suppose by contradiction that there exists a multiset $Q_{i^*}$ that occurs twice or more inside $x$.

First, we show that all the occurrences of $Q_{i^*}$ form a *run*, that is, any two such occurrences must overlap and there is no occurrence of $Q_k$ ($k \neq i^*$) between them. This is easy to see, since each $d_{i^*j}$ occurs only once in $x$.

Second, consider all the runs in $x$, where a multiset occurring once is seen as a degenerate run. If two runs intersect, their intersection contains exactly two symbols by Lemma 14.

Third, the run of $Q_{i^*}$ must be degenerate, thus contradicting the hypothesis that there are at least two occurrences. Indeed, if the run of $Q_{i^*}$ is not degenerate, then $|x| > |R|$, which is not possible. To see why, we recall that a valid solution $x$ of $\langle R, F \rangle$ is required to have size $|x| = |R| = 4|E| + 4$. Since $q = |\bigcup_i Q_i| = 4|E| + 2|V|$, some overlaps between consecutive runs are required. As previously mentioned, the intersection of two consecutive runs contains two elements. Hence, $r = 2|V| - 2$ is the number of symbols in the overlaps between pairs of consecutive runs in $x$. In order to fit the required length $|R|$, the first run must also intersect $R_w$ in $c_w$, while the

last one must intersect $R_s$ in $c_s$. We also should add to these $q$ elements, the two special symbols $c'_w$ and $c'_s$, totalizing $|x| = |R| = (q+2) - r$ elements in $x$ (and so many in $R$ as well). If the run of $Q_{i^*}$ is non-degenerate, then its size will be at least $|Q_{i^*}| + 1$, implying that there are at least $(q+2+1) - r > |R|$ symbols in $x$. Consequently, $|x|$ would be strictly larger than $|R|$, contradicting the validity of $x$ as solution of $\langle R, F \rangle$. □

It remains to prove Lemma 11 and Lemma 12.

Let us discuss Lemma 11. Given a Hamiltonian path $H = \langle i_1, i_2, \ldots, i_{|V|} \rangle$ of $G$, where $i_1 = w$ and $i_{|V|} = s$, in order to construct a solution $x$ of the corresponding #FMO instance $\langle R, F \rangle$, we arrange the multisets $Q_i$ in the same order as the corresponding vertices in $H$, as shown in Figure 5. The first symbol of $x$ is $c'_w$ and the last one is $c'_s$. Between them, $Q_{i_1}, Q_{i_1}, \ldots, Q_{i_{|V|}}$ appears in $x$, where the first symbol of $Q_{i_1}$ is $c_w$, and the last symbol is $i_1$, and the first symbol of $Q_{i_{|V|}}$ is $i_{|V|}$ and the last symbol is $c_s$. For the remaining $Q_{i_l}$, the first three symbols are $i_l, i_{l-1}$, and $d_{i_l i_{l-1}}$, and the first two of them overlap with $Q_{i_{l-1}}$ by Lemma 14. Analogously, the last three symbols are $d_{i_l i_{l+1}}, i_{l+1}$ and $i_l$, and the last two of them overlap with $Q_{i_{l+1}}$. The remaining symbols in $Q_{i_l}$ are $d_{i_l j}, j$ for all edges $\{i_l, j\} \in E$, such that $j \neq i_{l-1}, i_{l+1}$.

Each multiset $Q_{i_l}$ intersects $Q_{i_{l+1}}$ in $\{i_l, i_{l+1}\} \in E$. Note that, since $H$ is a Hamiltonian path, the symbols belonging to the union of all the intersections are $R' = \bigcup_{i \neq w, s} \{i, i\} \cup \{w, s\}$.

To prove that $x$ is a solution of $\langle R, F \rangle$, note that $x$ contains each multiset $Q_i$, $R_w$, $R_s$ by construction. As for each $Q_{ij} = \{d_{ij}, j\}$, we observe that its occurrence is contained in the occurrence of $Q_i$ in $x$. Moreover, $x$ contains the multiset $R$ and $x$ has size $|R|$, since $x$ is drawn from the multiset $\bigcup_i Q_i \cup \{c'_w, c'_s\} \setminus R'$, that is exactly the way $R$ is defined in $\langle R, F \rangle$. The above discussion prove Lemma 11.

To prove Lemma 12, consider a string $x \in \Sigma_H$, and $x' \in \Sigma_{H'}$ where $H' = \langle i'_1, i'_2, \ldots, i'_{|V|} \rangle$. Since $H \neq H'$, they must differ in at least one position $l$ (i.e. $i_l \neq i'_l$). Let assume w.l.o.g. that $|Q_{i_l}| \leq |Q_{i'_l}|$, and select the position $k$ of the leftmost symbol $d_{i_l j} \in Q_{i_l}$ occurring in $x$ for some $j$. Since the order of the multisets in $x$ is the same as that of the vertices in the Hamiltonian paths, $Q_{i_l} \neq Q_{i'_l}$ (since $i_l \neq i'_l$). By construction of the multisets, we have $d_{ij} \notin Q_{i'_l}$, then the $k$th symbol in $x$ and $x'$ differs, thus proving the claim.

### 4.3 Reduction from #HAM to #FMO

The #FMO problem is clearly in #$\mathcal{P}$, since we can take a solution string $x$ as a certificate. Therefore, we focus on its completeness.

We are given an undirected graph $G = \langle V, E \rangle$, along with its source $w$ and its destination $s$. The reduction goes as follows.

- Build an instance $\langle R, F \rangle$ as described in Section 4.1.

- Let $z$ be the number of solutions for the instance $\langle R, F \rangle$.

- Let $a = \prod_{i=1}^{|V|} \alpha_i \neq 0$, where $\alpha_i$ is defined as follows for a vertex $i$ of degree $d_i$:

$$\alpha_i = \begin{cases} 2^{(d_i - 1)} (d_i - 1)! & i = w, s \\ 2^{(d_i - 2)} (d_i - 2)! & i \neq w, s \end{cases}$$

- Return the integer $z/a$.

The above reduction takes polynomial time. To see its correctness, it suffices to show that $|\Sigma_H| = a$ for every Hamiltonian path $H = \langle i_1, i_2, \ldots, i_{|V|} \rangle$ in $G$.

We already proved in Section 4.2 that each solution $x \in \Sigma_H$ has the form reported in Figure 5. Here, the occurrence of each $Q_i$ is a sequence of pairs $Q_{ij} = \{d_{ij}, j\}$ except the first and the last symbol of $Q_i$. If $i \neq w, s$, the first and the last pairs always stay the same, while the remaining $d_i - 2$ pairs can be permuted in $(d_i - 2)!$ ways. For each such a way, we can permute each pair internally, thus giving an extra factor of $2^{d_i - 2}$. If $i = w, s$, we have $d_1 - 1$ pairs that can be permuted, yielding $2^{(d_i - 1)}(d_i - 1)!$ permutations.

**Theorem 16** *#FMO is #$\mathcal{P}$-complete.*

**Corollary 17** *Testing the C1P on multisets is $\mathcal{NP}$-complete.*

## 5 Conclusions

In this paper, we have shown that *counting* the number of orderings related to the C1P on multisets is #$\mathcal{P}$-complete. Hence, a polynomial-time algorithm is unlikely to exist, contrarily to what happens for sets. Although a direct mapping of the orderings for the C1P in multisets into the frontiers of PQ-trees has some intrinsic ambiguity, we proved that there exists an *indirect* mapping between the two counting problems. It would be interesting to find a direct and "natural" reduction between the two problems, without using the counting version of the Hamiltonian path as an intermediate problem (see Figure 2).